# Phase-Matched Second-Harmonic Generation in a flux grown KTP crystal ridge optical waveguide

VERONIQUE BOUTOU,[1,*] AUGUSTIN VERNAY,[1,*] CORINNE FELIX,[1] FLORENT BASSIGNOT,[2] MATHIEU CHAUVET,[3] DOMINIQUE LUPINSKI,[4] BENOIT BOULANGER,[1,**]

[1]Univ. Grenoble Alpes, CNRS, Institut Néel, 38000 Grenoble, France
[2]Femto-Engineering, 15B avenue des Montboucons, 25000 Besançon, France
[3] FEMTO-ST Institute, UMR CNRS 6174, Université de Franche-Comté, 15B avenue des Montboucons, 25000 Besançon, France
[4] Cristal Laser SA, Parc d'Activités du Breuil - 32, rue Robert Schuman 54850 Messein, France
*V. Boutou and A. Vernay contributed equally to this work.
**Corresponding author: benoit.boulanger@neel.cnrs.fr

**Type II second-harmonic generation was performed in a 15.8-mm-long KTiOPO$_4$ micrometric ridge waveguide with an average transversal section of 38 µm$^2$. Theoretical predictions are compared with experiments. Strong agreements are obtained for both phase-matching wavelengths and second-harmonic intensity. This work opens wide perspectives for integrated parametric optics.**

Because of its high optical nonlinearity [1], potassium titanyl phosphate KTiOPO$_4$ (KTP) is a crystal that has been extensively studied and is widely used in second-harmonic generation (SHG) or squeezed states devices for example [2]. In order to combine the large second-order nonlinear coefficient of KTP with a strong confinement in integrated photonics devices, there is a clear interest to elaborate waveguides with transverse micron-size dimensions. Among several waveguide-fabrication techniques such as proton exchange [3,4], ion implantation [5], laser micromachining [6] or etching on a platform [7,8] an alternative is the technique that consists in the manufactoring of ridge waveguides [9] with a high index contrast allowing a strong confinement of light in stable guiding structures. This method had been used earlier for instance in periodically poled LiNbO$_3$ ridge waveguides [10-12].

In the present study, we apply the ridge waveguide concept to KTP without any additional refractive index gradient [9] in order to maximize the light confinement. Another objective of this study is to show how the transversal dimensions of the waveguide modify the effective index and consequently the birefringence phase matching (BPM) conditions for SHG, which is a key parameter to define the targeted second-harmonic (SH) wavelength. We present a characterization of SHG in an x-cut KTP ridge waveguide. Prior to nonlinear experiments, we perform calculations for both x- and y-cut crystals. We further compare these calculations with the results of SHG experiments performed with a tunable coherent pump source. The SHG quantitative efficiency in intensity is also measured as a function of the incident intensity and compared with calculations.

Fabrication of the ridge waveguides begins with the cut of a flat rectangular sample from a KTP single crystal grown by the flux technique at Cristal Laser SA. The KTP dimensions are 20 mm along the x-axis, 10 mm along the y-axis, and 500 µm along the z-axis. One z-face is polished while all other faces are ground. A 300-nm-thick SiO$_2$ layer is deposited by ICPECVD onto the polished face followed by the sputtering of a 300-nm-thick gold layer [12]. A high flatness 3-inches diameter silicon wafer is also coated with a 300-nm-thick gold layer. The metallized faces of the crystal and the silicon wafer are brought in contact. This hybrid stacking is then pressed in an EVG wafer bonding machine and a high pressure is applied under vacuum at room temperature in order to prevent mechanical stress due to the dissimilar thermal coefficients [13]. A bonding of more than 98% of the surfaces in contact is measured by an ultrasound characterization technique. At this stage, a 1-mm-thick hybrid structure composed of a KTP sample bonded on a silicon substrate is obtained. The next step consists in thinning down the sample by grinding and polishing techniques to obtain a few µm-thick KTP layer. To this end, the KTP layer distribution is assessed by white-light optical reflectometry indicating a typical thickness variation of 1µm or less along x. In addition, the surface is analyzed by optical profilometry and atomic force microscopy [12]. The next step consists in using a precision dicing saw equipped with a 56-mm-diameter and 400-µm-thick diamond blades to define the ridges with low blade buckling in the hybrid stack [12]. Micrometric ridges with roughness sides of about 5 nm RMS are manufactured by adjusting the rotation speed and cutting speed of the dicing process to the material [15,16]. Because the KTP layer thickness is uneven when approaching the sample edge, the waveguides are cut transversely in the more uniform part to form 15.8-mm-long waveguides along the x-axis. Six waveguides were diced out of the initial KTP sample. Fig. 1(a) shows the image of a 7 x 5 µm² cross section waveguide. Side walls of the ridge are curved due the blunt corner of the blade. We selected the best waveguide among all in order to perform the experiments described hereafter.

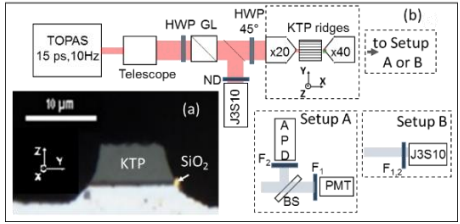

Fig. 1. (a) Image (MOx100) of the input section of a ridge with a cross section of 7 x 5 µm². (b) Experimental setup: HWP half wave plate ; GL Glan Taylor prism ; PMT photomultiplier module (Hamamatsu) ; APD InGaAs avalanche photodiode (Thorlabs) ; J3S10 joulemeter (Molectron) ; $F_1$ FGB37 filter to block the fundamental; $F_2$ long pass filter to block the SH ; ND: neutral density filter ; BS: beam-splitter.

Ideally, square section waveguides are favored since it facilitates excitation of the waveguide fundamental mode. A good overlap is obtained if butt-coupling is realized with a Gaussian beam of proper dimension [16]. The effective indices of the guided modes have to be determined in order to accurately calculate the SHG BPM wavelengths and to properly design the corresponding waveguides. This task is performed with the commercial software COMSOL based on the finite elements method. A square S = d x d waveguide structure with propagation along the x-axis of KTP is modeled. The bulk KTP refractive index is obtained from Sellmeier equations taken from the literature [17]. For a given waveguide dimension, the effective index of the fundamental guided mode is calculated for both z-polarized ($TM_0$) and y-polarized ($TE_0$) light, by considering wavelengths varying between 0.5 to 3.0 µm. Dispersion equations giving the effective indices are retrieved by fitting those numerical data assuming a Sellmeier-like form. This method is iterated for waveguide dimension d spanning from 3 to 10 µm. The best fitting equation we found, with a discrepancy below $10^{-4}$, is:

$$\left(n_{eff}^\omega\right)_i^d = \left(A_i^d \times \lambda_\omega^{B_i^d} + \frac{C_i^d}{10^{-6} \times \lambda_\omega^{D_i^d} - E_i^d} - F_d^i \times 10^{-6} \times \lambda^{G_i^d}\right)^{H_i^d} + I_d^i \quad (1)$$

with i=x,y,x.

Table 1 gives, as an example, the corresponding dispersion coefficients for a waveguide with d = 6 µm. Fig. 2 gives the evolution of the effective indices as a function of the waveguide size calculated at the wavelength λ = 1 µm according to Eq. (1). The effective indices decrease from the bulk value with the transversal dimension of the ridge, which will necessarily imply a change in the BPM conditions.

Note that the numerical Eq. (1) keeps some physical meaning since the quantity $\lambda_i^0 = \left(E_i^d \times 10^6\right)^{1/D_i^d}$, which should represent the wavelength corresponding to the ultraviolet oscillator frequency, is very close to that of bulk KTP, *i.e.*: 0.317 µm for the x-component from Table 1, compared to 0.206 µm from the Sellmeier equation of reference [17]. The proximity is of same kind for the y- and z-components of the effective index. In the following, we specifically discuss the type II BPM for two different orientations of the KTP ridge: the x-cut that has been investigated experimentally here, and the y-cut. We assumed that both fundamental and SH waves propagate in the ridge in their fundamental mode with a perfect overlap. For an x-cut (respectively y-cut) ridge, the energy conservation law is:

$$\frac{1}{\lambda_{2\omega}^{y,x}} = \frac{1}{\lambda_\omega^z} + \frac{1}{\lambda_\omega^{y,x}} \quad (2)$$

where x, y and z stand for the directions of polarization of the waves at the fundamental wavelength, $\lambda_\omega$, and second-harmonic wavelength, $\lambda_{2\omega}$. The corresponding mismatch relation $\Delta k(\omega, d)$ writes as:

$$\Delta k(\omega, d) = \frac{1}{c}\left[\omega\left(n_{eff}^\omega\right)_d^j + \omega\left(n_{eff}^\omega\right)_d^z - 2\omega\left(n_{eff}^{2\omega}\right)_d^j\right]$$

where
$$j = \begin{cases} x \text{ for a } y-cut \text{ ridge} \\ y \text{ for a } x-cut \text{ ridge} \end{cases} \quad (3)$$

In Fig. 3, the BPM fundamental wavelength is calculated assuming $\Delta k(\omega_{PM}, d) = 0$ as a function of d. For both x-cut and y-cut KTP ridges, the BPM fundamental wavelength drastically increases when d decreases. This is the signature of a size effect of the waveguide. The bulk limits are indicated by a horizontal dashed line depending on the orientation axis of the cut. It is then clear that the transversal dimension of the guide can be used to efficiently tune the BPM wavelength: as an example, an y-cut KTP ridge with d = 4.6 µm (see $Y_1$ in Fig. 3) allows SHG to be phase-matched at a fundamental wavelength of 1.064 µm, the well-known fundamental wavelength of Nd-YAG laser. We will further take benefit of these calculations for SH experiments in an x-cut KTP ridge.

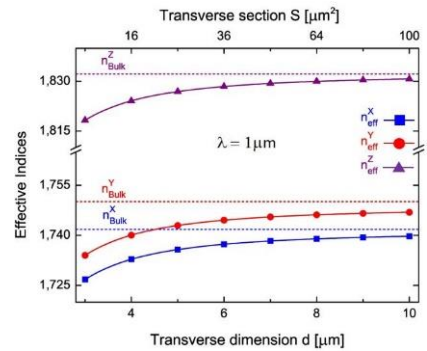

Fig. 2. Effective refractive indices of KTP ridges at λ= 1 µm calculated as a function of the transversal dimension d or cross section S = d x d

**Table 1. Dispersion coefficients for a waveguide transverse dimension d = 6μm.**

| i | $A_i(nm^{-1})$ | $B_i$ | $C_i(nm)$ | $D_i$ | $E_i(nm)$ | $F_i(nm^{-1})$ | $G_i$ | $H_i$ | $I_i$ |
|---|---|---|---|---|---|---|---|---|---|
| x | 0.4488 | 0.01948 | 0.5288 | 2.277 | 0.4939 | 0.88 | 1.615 | 0.07128 | 0.7808 |
| y | 1.017 | 0.004481 | 1.47 | 2.323 | 0.6407 | 1.242 | 1.653 | 0.07207 | 0.7373 |
| z | 2.236 | 0.006506 | 0.8399 | 2.14 | 0.1814 | 0.5619 | 1.811 | 0.1125 | 0.7183 |

A scheme of the SH experimental setup is given in Fig. 1(b). The output of a tunable optical parametric amplifier (TOPAS, repetition rate 10 Hz pulse duration FWHM 15 s, linear polarization). In order to control and adapt the fundamental intensity, a combination of a half-wave plate (HWP) and a Glan-Taylor prism is installed. A second HWP allows the polarization of the fundamental to be fixed at 45° with respect to the y- and z-axes of the input face. A x20 microscope objective (MO) is used to inject the incident pump beam with the right polarization on each of the KTP ridges. The sample is placed on a 3-axes micrometric translation stage in order to properly adjust the injection of the pump beam in one of the ridge. At the output of the ridge, the residual transmitted fundamental beam and the SH beam are collected with a x40 MO also mounted on a 3-axes translation stage. Two types of experiments are performed using setup A and setup B respectively.

The first experiment (Fig. 1(b) setup A) is devoted to the wavelength dependence of the SH signal. The pump beam is tuned between 1.05 μm and 1.20 μm. At the output of the ridge, a beam-splitter is used to separate the transmitted pump and the SH. The fundamental beam is collected by an avalanche InGaAs photodiode (APD) whereas a photomultiplier module (PMT) is chosen to collect the SH signal. Additional filters $F_1$ in front of the PMT module and $F_2$ in front of the APD are used to prevent any cross talking on both fundamental and SH signals.

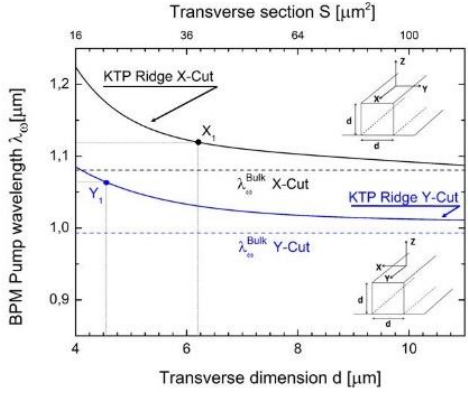

Fig. 3. Calculated SHG BPM fundamental wavelength $\lambda_\omega$ *versus* the transversal dimension d of an x-cut (top curve) and an y-cut (bottom curve). The dashed lines indicate the bulk limit for an x-cut (top) and y-cut (bottom) KTP.

Figure 4 gives the result of the SHG wavelength dependence that we obtained. For this experiment, the incident pump beam intensity is estimated to be $I_{pump} = 0.47$ GW/cm². The BPM wavelength is measured to be $\lambda_\omega^{PM} = 1.120$ μm. A comparison with the calculations (Fig. 3, data point $X_1$) for x-cut ridges indicates that $\lambda_\omega^{PM}$ is typical of a ridge with a section of about $S_{BPM} = 38$ μm², *i.e.* d close to 6.2 μm. $S_{BPM}$ has to be compared with the input and output sections that we have measured from profiles similar to Fig 1(a).

We obtained respectively $S_{in} = 43$ μm² and $S_{out} = 35$ μm², showing that $S_{BPM}$ belongs to this interval. This experiment well validates our calculation of the effective refractive indices. The agreement, together with the output beam profiles that filled the entire output section as we have observed on a camera, also confirms that the fundamental and SH waves propagate in the fundamental mode of the waveguide.

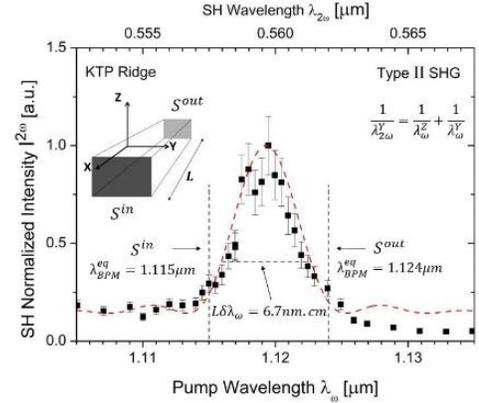

Fig. 4. SH normalized intensity versus the fundamental wavelength obtained for a waveguide with a length L = 15.8-mm-long. The black squares correspond to the experimental data. The dashed curve is a fit by a *sinc²* function, from which we deduced the wavelength acceptance $L.\delta\lambda_\omega$. The two vertical dashed lines indicate the BPM wavelength that should correspond to the measured input ($S_{in}=43$ μm²) and output ($S_{out}=35$ μm²) section of the waveguide.

From this fit it is possible to retrieve a wavelength acceptance of the BPM SHG equal to $L.\delta\lambda_\omega = 6.7$ nm.cm, with L the length of the ridge. This value is one order of magnitude larger than the 0.65 nm.cm that are expected for a uniform ridge. This discrepancy can be mainly attributed to manufacturing defects formed during the polishing step or at the dicing stage. Indeed, the section of the guide is not constant from the entrance ($S_{in}$) to the exit ($S_{out}$), leading to a variation of the BPM wavelength along the direction of propagation, *i.e.* the x-axis. As stated in Fig. 4 (dashed vertical lines) and with the help of calculations in Fig. 3, we attribute an equivalent BPM wavelength of respectively $\lambda_{BPM}^{eq} = 1.115$ μm for $S_{in}$ and $\lambda_{BPM}^{eq} = 1.124$ μm for $S_{out}$, which roughly corresponds to the wavelength acceptance that is observed.

The second experiment (Fig. 1(b) setup B) is dedicated to the absolute measurement of the SH intensity as a function of the fundamental beam intensity injected in the ridge. For that purpose, the fundamental wavelength is fixed at the BPM value $\lambda_\omega^{PM}=1.120$ μm found in the first experiment. The pump intensity is then changed and monitored prior to the injection in the ridge with a J3S10 joulemeter using additional neutral density filters whereas the SH signal is simultaneously monitored at the output of the ridge on a second J3S10 detector. In Fig. 5, the intensity of the pump is tuned between 0.2 and 22.5 GW/cm$^2$, while the corresponding SH intensity is measured. These experimental results have been compared with calculation. In the general case, *i.e.* when the pump is depleted, the analytical solution of the coupled equations for SHG is easily accessible in the parallel beam approximation and by neglecting the spatial walk-off effect. The SHG intensity under phase-matching is then given by [18]:

$$I^{2\omega}(L) = I^{\omega}(0) T^{\omega} T^{2\omega} th^2 \left( \frac{\omega d_{eff}}{c n_{eff}^{2\omega}} \sqrt{\frac{2 T^{\omega} I^{\omega}(0)}{\varepsilon_0 c n_{eff}^{\omega}}} L \right) \quad \textbf{(4)}$$

$I^{\omega}(0)$ is the input fundamental intensity, *i.e.* just before the waveguide entrance, $T^{\omega}$ and $T^{2\omega}$ are respectively the fundamental and SH transmission coefficients, and $d_{eff} = 2.59\ pm.V^{-1}$ is the second-order effective coefficient calculated at $\lambda_{\omega}^{BPM}$ obtained from [1]. We assumed its value to be the same as in the bulk according to our fabrication process. L is the propagation length. The effective indices are taken at d = 6.2 µm, which corresponds $\lambda_{\omega}^{BPM}$ =1.120 µm according to Fig. 4.

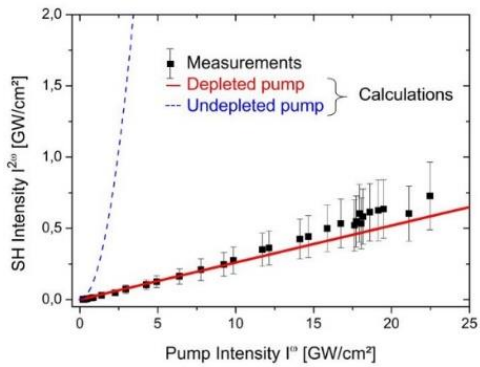

Fig. 5. Output SH intensity as a function of the input fundamental beam intensity at $\lambda_{\omega}^{BPM}$ = 1.120 µm. The black squares correspond to the experimental data, the error bars being estimated for each pump intensity taking into account the standard deviations on each signal strength averaged on a scope.

We estimated $T^{\omega}$ and $T^{2\omega}$ by measuring the intensity incident on the waveguide and at its exit of two continuous laser beams at 1.064 µm and 0.532 µm. For the KTP guide itself, we measured $T^{\omega}_{meas.}$= 7.5% and $T^{2\omega}_{meas}$ = 18.5% but a further correction was added to remove the coupling efficiency (Fresnel contribution and mode matching) of the 1.064 µm beam at the exit of the waveguide, and that of the beam at 0.532 µm at the entrance, which leads to $T^{\omega}$= 9.3% and $T^{2\omega}$= 28%. Using these values in Eq. (4) without any fitting parameter, we highlight a very good agreement between the calculated and experimental data in the case of pump depletion (see Fig 5). A maximum intensity SHG conversion efficiency of 3.4 % is measured, lower than that in reference [9] because of weak input and output coupling efficiencies. However, the depleted pump regime is achieved, which proves that an efficient process exists. Moreover, besides internal propagation losses estimated to be 0.6 dB/cm (1.6 dB/cm) at the pump (SH) wavelengths, our waveguides are very robust. In future generations, the sidewalls and top roughness of the samples, as well as the input and output faces optical qualities, will need to be improved in order to reduce losses and to increase the confinement [12]. The SiO$_2$ layer thickness has also to be optimized. Further experiments will be then necessary to compare the robustness of the elaboration procedure in regards with other techniques [4,9]. Indeed, the optimal conversion efficiency we calculated with Fresnel as only source of losses is 85.3%, which gives us the margin of progress of our elaborating process.

As a conclusion, we fabricated ridge waveguides carved in a KTP single crystal. We determined the corresponding wavelength dispersion equations of the effective refractive indices that can be derived for any transversal dimension of the waveguide. We were subsequently able to accurately retrieve the type II SHG BPM conditions that were significantly different than those in the bulk material, which highlights the fact that tailoring the waveguide transversal dimension can be used to reach BPM for specific wavelengths. Moreover, the theoretical prediction of SHG conversion efficiency fully matched the experimental results, and that without any fitting parameters. Pump depletion has been achieved. Higher quality sample and extending the waveguide length could lead to even better results, which opens new field for integration of KTP waveguides in photonic nonlinear devices.

**Acknowledgment**. Institut Néel would like to acknowledge David Jegouso for his technical support. This work is supported by the Agence Nationale de la Recherche through the project TRIQUI (ANR-17-CE24-0041-03), and partly by the RENATECH network and its FEMTO-ST MIMENTO technological facility.

**References**

1. B. Boulanger, J.P. Fève, P. Delarue, I. Rousseau, and G. Marnier, J. Phys. B At. Mol. Opt. Phys. **32**, 475 (1999).
2. J. Gao, F. Cui, C. Xue, C. Xie, and P. Kunchi, Opt. Lett. **23**, 870 (1998).
3. K.R. Parameswaran, R.K. Route, J.R. Kurz, R. V. Roussev, M.M. Fejer, and M. Fujimura, Opt. Lett. **27**, 179 (2002).
4. W. M. F. Volk, C. E. Rüter and D. Kip, SPIE LASE, 2018.
5. J.-H. Zhao, X.-F. Qin, F.-X. Wang, G. Fu, J. Du, and X.-L. Wang, Opt. Mater. Express **3**, 954 (2013).
6. L. Li, W. Nie, Z. Li, Q. Lu, C. Romero, J.R. Vázquez De Aldana, and F. Chen, Sci. Rep. **7**, 1 (2017).
7. X. L. Chang, Y. Li, N. Volet, L. Wang, J. Peters and J. E. Bowers, Optica, **3**, 531 (2016).
8. Y. C. Wang, X. Xiong, N. Andrade, V. Venkataraman, X-F. Ren, G-C. Guo and M. Lončar, Optics Express, **25**, 6963 (2017).
9. C. Chen, C.E. Rüter, M.F. Volk, C. Chen, Z. Shang, Q. Lu, S. Akhmadaliev, S. Zhou, F. Chen, and D. Kip, Opt. Express **24**, 16434 (2016).
10. R. Kou, S. Kurimura, K. Kikuchi, A. Terasaki, H. Nakajima, K. Kondou, and J. Ichikawa, Opt. Express **19**, 11867 (2011).
11. S. Kurimura, Y. Kato, M. Maruyama, Y. Usui, and H. Nakajima, Appl. Phys. Lett. **89**, 191123 (2006).
12. M. Chauvet, F. Henrot, F. Bassignot, F. Devaux, L. Gauthier-Manuel, V. Pecheur, H. Maillotte, and B. Dahmani, J. Opt. U.K. **18**, 1 (2016).
13. J.D. Bierlein and H. Vanherzeele, J. Opt. Soc. Am. B **6**, 622 (1989).
14. M.F. Volk, S. Suntsov, C.E. Rüter, and D. Kip, Opt. Express 24, 1386 (2016).
15. N. Courjal, B. Guichardaz, G. Ulliac, J.Y. Rauch, B. Sadani, H.H. Lu, and M.P. Bernal, J. Phys. D. Appl. Phys. **44**, 305101 (2011).
16. F. Devaux, E. Lantz, and M. Chauvet, J. Opt. Soc. Am. B **33**, 703 (2016).
17. K. Kato, J. Quantum Electron. **27,** 1137 (1991).
18. R. Eckardt and J. Reintjes, J. Quantum Electron. **20**, 1178 (1984)